%Paper: hep-ph/9208221
%From: manohar@sphal.ucsd.edu
%Date: Tue, 11 Aug 92 14:07:03 PDT

%%%%
%
\input harvmac
%%%%%%%%%%%%%%%%%%%%%%%%%%%%%%%%%%%%%%%%%%%%%%%%%%%%%%%%%%%%%%%%%%%%%%
%
%  UCSD macros to overwrite some of the definitions in harvmac.tex
%  (include after harvmac.tex)
%  last modified 4/92
%
%%%%%%%%%%%%%%%%%%%%%%%%%%%%%%%%%%%%%%%%%%%%%%%%%%%%%%%%%%%%%%%%%%%%%%%
%
% modify the output routine for the little format
%
\ifx\answ\bigans
\else
\output={
  \almostshipout{\leftline{\vbox{\pagebody\makefootline}}}\advancepageno
}
\fi
%
%
% address
%
\def\mayer{\vbox{\sl\centerline{Department of Physics 0319}%
\centerline{University of California, San Diego}
\centerline{9500 Gilman Drive}
\centerline{La Jolla, CA 92093-0319}}}
%
% grant numbers
%
\def\doe{\#DOE-FG03-90ER40546}
\def\pyiam{PHY-8958081}

%
% preprint number
%
\def\UCSD#1#2{\noindent#1\hfill #2%
\bigskip\supereject\global\hsize=\hsbody%
\footline={\hss\tenrm\folio\hss}}% restores pagenumbers
%
% abstract
%
\def\abstract#1{\centerline{\bf Abstract}\nobreak\medskip\nobreak\par #1}
%
%
% titlefont
%
%
\edef\tfontsize{ scaled\magstep3}
 \tfontsize  \tfontsize
 \tfontsize \font\titlei=cmmi10 \tfontsize
\font\titleis=cmmi7 \tfontsize \font\titleiss=cmmi5 \tfontsize
\font\titlesy=cmsy10 \tfontsize \font\titlesys=cmsy7 \tfontsize
\font\titlesyss=cmsy5 \tfontsize  \tfontsize
\skewchar\titlei='177 \skewchar\titleis='177 \skewchar\titleiss='177
\skewchar\titlesy='60 \skewchar\titlesys='60 \skewchar\titlesyss='60
%
%\def\titlefont{\def\rm{\fam0\titlerm}% switch to title font
%\textfont0=\titlerm \scriptfont0=\titlerms \scriptscriptfont0=\titlermss
%\textfont1=\titlei \scriptfont1=\titleis \scriptscriptfont1=\titleiss
%\textfont2=\titlesy \scriptfont2=\titlesys \scriptscriptfont2=\titlesyss
%\textfont\itfam=\titleit \def\it{\fam\itfam\titleit}\rm}
%
%
% math symbols
%
%---------------------------------------------------------------------
%
\def\inv{^{\raise.15ex\hbox{${\scriptscriptstyle -}$}\kern-.05em 1}}
  %prime
\def\lbar{{\lower.35ex\hbox{$\mathchar'26$}\mkern-10mu\lambda}} %lambda bar

%
%
% various slashed symbols
%
%
\def\slash#1{\rlap{$#1$}/} % slashes a character
\def\dsl{\,\raise.15ex\hbox{/}\mkern-13.5mu D} %this one can be subscripted
\def\delsl{\raise.15ex\hbox{/}\kern-.57em\partial}
\def\Ksl{\hbox{/\kern-.6000em\rm K}}
\def\Asl{\hbox{/\kern-.6500em \rm A}}
\def\Dsl{\hbox{/\kern-.6000em\rm D}} %roman D
\def\Qsl{\hbox{/\kern-.6000em\rm Q}}
\def\gradsl{\hbox{/\kern-.6500em$\nabla$}}
%
% space and backspace in l mode
%
\def\lspace{\ifx\answ\bigans{}\else\qquad\fi}
\def\lbspace{\ifx\answ\bigans{}\else\hskip-.2in\fi} % $$\lbspace...$$
%
%     boxes an equation
%
\def\boxeqn#1{\vcenter{\vbox{\hrule\hbox{\vrule\kern3pt\vbox{\kern3pt
        \hbox{${\displaystyle #1}$}\kern3pt}\kern3pt\vrule}\hrule}}}
%
%     draw a little box (end of proof symbol)
%     e.g. \mbox{.1}{.1}
%
\def\mbox#1#2{\vcenter{\hrule \hbox{\vrule height#2in
\kern#1in \vrule} \hrule}}
%
%
%
%     curly letters
%
   %curly letters

 \def\CR{{\cal R}}

%
%
%
%     derivatives
%
%

%

\def\bar#1{\overline{#1}}

\def\bra#1{\left\langle #1\right|}
\def\ket#1{\left| #1\right\rangle}
\def\abs#1{\left| #1\right|}

\def\darr#1{\raise1.5ex\hbox{$\leftrightarrow$}\mkern-16.5mu #1}

%
 %pound sterling
%
\def\half{{\textstyle{1\over2}}} %puts a small half in a displayed eqn
\def\frac#1#2{{\textstyle{#1\over #2}}} %puts a small fraction
%in a displayed eqn
%
%
%     various math operators
%
%

\def\Tr{\mathop{\rm Tr}}

\def\MeV{{\rm MeV}}

%
%
%
%

%
%       relations
%
\def\ltap{\ \raise.3ex\hbox{$<$\kern-.75em\lower1ex\hbox{$\sim$}}\ }
\def\gtap{\ \raise.3ex\hbox{$>$\kern-.75em\lower1ex\hbox{$\sim$}}\ }
\def\gl{\ \raise.5ex\hbox{$>$}\kern-.8em\lower.5ex\hbox{$<$}\ }
\def\roughly#1{\raise.3ex\hbox{$#1$\kern-.75em\lower1ex\hbox{$\sim$}}}
%
%
%       This defines et al., i.e., e.g., cf., etc.
\def\ie{\hbox{\it i.e.}}        \def\etc{\hbox{\it etc.}}
\def\eg{\hbox{\it e.g.}}        
\def\etal{\hbox{\it et al.}}

\def\np#1#2#3{{Nucl. Phys. } B{#1} (#2) #3}
\def\pl#1#2#3{{Phys. Lett. } {#1}B (#2) #3}
\def\prl#1#2#3{{Phys. Rev. Lett. } {#1} (#2) #3}
\def\physrev#1#2#3{{Phys. Rev. } {#1} (#2) #3}

\relax

\noblackbox

\def\twolr{SU(2)_L\times SU(2)_R}
\def\threelr{SU(3)_L\times SU(3)_R}
\def\dim{{\rm dim\,}}
\def\kket#1{\left.\left|#1\right\rangle\!\right\rangle}
\def\bbra#1{\left\langle\!\left\langle#1\right.\right.}
\def\sket#1{\left|#1\right)}
\def\hket#1{\left|#1\right\}}
\def\sbra#1{\left(#1\right|}
\def\hbra#1{\left\{#1\right|}
\def\clebsch#1#2#3#4#5#6{\left(\left.
\matrix{#1&#2\cr#4&#5\cr}\right|\matrix{#3\cr#6}\right)}
\def\smallclebsch#1#2#3#4#5#6{\left(\left.
\matrix{#1&#2\cr\noalign{\smallskip}#4&#5\cr}%
\right|\matrix{#3\cr\noalign{\smallskip}#6}\right)}

\centerline{{\titlefont{Properties of Baryons Containing a Heavy Quark}}}
\medskip
\centerline{{\titlefont{in the Skyrme Model}}}
\bigskip
\centerline{Zachary Guralnik, Michael Luke and Aneesh V. Manohar}
\bigskip
\mayer
\vfill
\abstract{The properties of baryons containing one heavy quark are
studied in the Skyrme model,
where they are treated as bound states of a
heavy meson with an $SU(2)$ chiral soliton.
In the large $N_c$ limit,
the baryon spectrum is an infinite tower of degenerate
states with isospin and spin of the light degrees of freedom
$I=s_\ell=0,1,2,\dots$, and total spin $s=\abs{s_\ell\pm1/2}$.
Exotic states with no quark model
analogues in the large $N_c$ limit are found to be unbound.  The
$\Sigma_c-\Lambda_c$
mass difference is computed and found to be in good agreement with
experiment.  The $\Sigma_c\,\Sigma_c\,\pi$ and $\Sigma_c\,\Lambda_c\,\pi$
coupling constants are calculated to leading order in $N_c$ in terms of
$g_A$, the axial coupling constant of the nucleon.
We discuss the extension of our results to $SU(3)$ chiral solitons.
}
\vfill
%\draftmode
\UCSD{\vbox{\hbox{UCSD/PTH 92-24}\hbox{hep-ph/9208221}}}{August 1992}

\newsec{Introduction}
Skyrme originally suggested that baryons are solitons in the non-linear chiral
lagrangian used to describe the self-interactions of the Goldstone pions
\ref\skyrme{T.H.R. Skyrme, Proc. Roy. Soc. {A260} (1961) 127}.
The solitons of the chiral lagrangian have the correct quantum numbers to be
the baryons \ref\witten{E. Witten, \np{223}{1983}{433}} provided one includes
the Wess-Zumino term \ref\wesszumino{J. Wess and B. Zumino, \pl{37}{1971}{95}}.
The model of QCD baryons as solitons can be used to compute many of their
properties \ref\anw{G.S. Adkins, C.R. Nappi and E. Witten,
\np{228}{1983}{552}}. The soliton model was originally considered in Ref.~\anw\
for the case of two light flavours, but can be generalised to three light
flavours in two different ways. The first method is to treat the $s$ quark as
light, and consider the solitons of the $\threelr$ chiral lagrangian
\ref\guad{E. Guadagnini, \np{236}{1984}{35}}\ref\am{A.V. Manohar,
\np{248}{1984}{19}}. The second method is to treat the strange quark as heavy,
and consider baryons containing a $s$ quark as soliton-$K$~meson bound states
\ref\calkleb{C. Callan and I. Klebanov, \np{262}{1985}{365},
\pl{202}{1988}{260}}, where the solitons are those of an $\twolr$ chiral
lagrangian. Heavy quarks such as the $c$ and $b$ quarks have masses which are
not small compared with the scale of chiral symmetry breaking. Thus baryons
containing a heavy quark can only be treated by the second method, as bound
states of solitons with heavy $D$ and $B$ mesons \ref\mrho{M. Rho, D.O.
Riska, and N.N. Scoccola, \pl{251}{1990}{597},
Z.~Phys. {A341} (1992) 343\semi
Y. Oh, D. Min, M. Rho, and N. Scoccola, Nucl. Phys. A534 (1991)
493}\ref\jmw{E. Jenkins, A.V.
Manohar and M.B. Wise, Caltech Preprint CALT-68-1783 (1992)}.
Callan and Klebanov considered a theory with an $\twolr$ chiral
lagrangian as well as $K$ mesons by expanding an $\threelr$ chiral lagrangian,
so that the $K$ meson couplings to the pions were determined by chiral
$\threelr$ symmetry. They also neglected the $K^*$ meson because it is much
heavier than the $K$ meson. While these approximations are reasonable for the
$s$ quark, they are not valid for the $c$ and $b$ quarks. A better starting
point for heavy quarks is to expand about the limit $m_Q\rightarrow\infty$. In
this limit, the pseudoscalar and vector mesons, such as the $D$ and $D^*$ (or
$B$ and $B^*$) are degenerate, and must {\sl both} be included in the chiral
lagrangian as matter fields. In addition, the couplings of the heavy mesons are
no longer related to the pion couplings by chiral symmetry transformations
involving the heavy quark. The analysis of baryons containing a heavy
quark in Ref.~\mrho\ was done without including $D^*$ and $B^*$ mesons,
and hence does not respect the heavy quark symmetries.

We first review the formalism for heavy-meson--soliton bound states of
Ref.~\jmw\ that is necessary for the computations discussed in this paper. We
then compute the spectrum of baryons containing a heavy quark in the case of
two light flavours, using the soliton solution of the $\twolr$ chiral
lagrangian. In Sec.~4, the $\Sigma_c-\Lambda_c$ mass difference is computed in
terms of the $\Delta-N$ mass difference. The result agrees well with
experiment. In Sec.~5, we discuss the quantum number $K$ which can be used to
label the soliton--heavy-meson bound states. The energies of the bound states
depend only on $K$ in the large $N_c$ limit, which explains the degeneracy of
the spectrum computed in Sec.~3. In Sec.~6, we compute the pion-baryon coupling
constants in terms of the $D$-meson--pion coupling constant
and $g_A$, the axial
vector coupling of the nucleon. In Sec.~7, we generalise the results of the
previous sections to the $SU(3)$ case.

\newsec{The Formalism for Studying Soliton-Meson Bound States}
The basic formalism for studying properties of baryons containing a heavy quark
in the soliton framework was described in detail in \jmw. In the limit
$m_Q\rightarrow\infty$, QCD has a heavy quark spin-flavour symmetry that
determines many of the properties of hadrons containing a heavy quark
\ref\wisgur{N. Isgur and M.B. Wise, \pl{232}{1989}{113},
\pl{237}{1990}{527}}\ref\georgi{H. Georgi, \pl{240}{1990}{447}}. In the heavy
quark limit, the total angular momentum of the light degrees of freedom (light
quarks and gluons) $\vec S_\ell$ is conserved, where
\eqn\lights{
\vec S_\ell = \vec S - \vec S_Q,
}
$\vec S$ is the total spin, and $\vec S_Q$ is the spin of the heavy quark.
The lowest lying mesons containing a heavy quark with $Q\bar q_a$ ($q_1=u$,
$q_2=d$, $q_3=s$) flavour quantum numbers have $s_\ell$, the eigenvalue of
$S_\ell^2=s_\ell(s_\ell+1)$ equal to 1/2, and come in a degenerate doublet
containing a spin zero pseudoscalar and a spin one vector. For $Q=c$, these are
the $D$ and $D^*$ mesons. The heavy meson field for the ground state $Q\bar
q_a$ mesons is written as a $4\times 4$ bispinor matrix,
\eqn\hmatrix{
H_a = {(1 + \slash v)\over 2} \left[ P^*_{a\mu}\gamma^\mu - P_a
\gamma_5\right],
}
where $v^\mu$ is the heavy quark four-velocity, $v^2=1$. The fields $P_a$ and
$P^*_{a\mu}$ destroy the heavy pseudoscalar and vector particles that comprise
the ground state $s_\ell=1/2$ doublet, and satisfy the constraint
$P^*_{a\mu}v^{\mu}=0$. The transformation rule for the $H$ field under
$SU(2)_Q$ heavy quark spin symmetry is
\eqn\hsymmetry{
H_a\rightarrow S H_a,
}
where $S\in SU(2)_Q$, and the transformation rule under chiral $\threelr$ is
\eqn\hchiral{
H_a\rightarrow \left( H R^\dagger\right)_a,
}
where we will use the primed basis for the $H$ fields defined in Ref.~\jmw.
It is also convenient to define
\eqn\hbardef{
\bar H^a = \gamma^0 H_a^\dagger \gamma^0 =
\left[ P_{a\mu}^{*\dagger}\gamma^\mu + P_a^\dagger \gamma_5\right]
{(1 + \slash v)\over 2}.
}
The Goldstone boson fields have
the $\threelr$ transformation law
\eqn\sigmatrans{
\Sigma(x) \rightarrow L\ \Sigma(x)\ R^\dagger.
}
The $\Sigma$ field can be written in terms of the pion fields as
\eqn\sigmapion{
\Sigma(x) = e^{2i M /f},
}
where $M$ is the Goldstone-boson matrix
\eqn\Mdef{
M = \left[\matrix{
\pi^0/\sqrt2+\eta^0/\sqrt6 & \pi^+ & K^+ \cr
\noalign{\smallskip}
\pi^- & -\pi^0/\sqrt2+\eta^0/\sqrt 6 & K^0 \cr
\noalign{\smallskip}
K^- & \bar K^0 & - 2 \eta^0/\sqrt 6 \cr
}\right],
}
and the pion decay constant $f\approx 132$~MeV. The transformation rules under
parity are \jmw\
\eqn\hparity{
H_a(x^0,\vec x) \rightarrow \gamma^0 H_b(x^0,-\vec x) \gamma^0 \Sigma^{\dagger
b}{}_a(x^0,-\vec x),
}
and
\eqn\sigmaparity{
\Sigma(x^0,\vec x)\rightarrow \Sigma^\dagger(x^0,-\vec x).
}

The chiral lagrangian density for heavy-meson--pion interactions
is \ref\wise{M.B. Wise, \physrev{D45}{1992}{2118}} \ref\burdman{G. Burdman and
J.F. Donoghue, \pl{280}{1992}{287}} \ref\yan{Tung-Mow Yan, Hai-Yang Cheng,
Chi-Yee Cheung, Guey-Lin Lin,
Y.C. Lin and Hoi-Lai Yu, CLNS 92/1138 (1992)}
\eqn\intlag{\eqalign{
{\cal L} &= - i \Tr \bar H v_\mu \partial^\mu H
+ {i\over 2} \Tr
\bar H H v^\mu (\Sigma^{\dagger} \partial_\mu \Sigma)\cr
&+ {ig\over 2} \Tr \bar H H \gamma^\nu \gamma^5
(\Sigma^{\dagger} \partial_\nu \Sigma) + \ldots ,
}}
where the ellipsis denotes terms with more derivatives and the trace is over
spinor and flavour indices.  Eq.~\intlag\ is
the most general lagrangian density
invariant under chiral $\threelr$,
heavy quark spin symmetry and parity.  It is easy to
generalise this lagrangian density to include
explicit $\threelr$ symmetry breaking from $u$, $d$, and $s$
quark masses and explicit $SU(2)_Q$ symmetry breaking from
$\Lambda_{QCD}/m_Q$ effects.
The coupling $g$ determines the $D^* \rightarrow D\pi$ decay width,
\eqn\width{\Gamma (D^{*+} \rightarrow D^0 \pi^+) = {1\over 6\pi} ~{g^2\over
f^2} |\vec p_\pi|^3 \,\, .
}
The present experimental limit \ref\accmor{The ACCMOR Collaboration (S. Barlag,
\etal), \pl{278}{1992}{480}}, $\Gamma (D^{*+} \rightarrow D^0
\pi^+)\ltap\ 72\ {\rm keV}$, implies that $g^2  \ltap 0.4$.

\newsec{Masses of Baryons containing a Heavy Quark: The $SU(2)$ Case}

The masses of baryons containing a heavy quark are computed in this section for
the case of two light flavours.
The algebra is considerably simpler than for the
$SU(3)$ case, which is studied in Sec.~7. We will use the fields and
interaction lagrangian of the previous section, with the light quark index
restricted to the values 1,2.

The soliton solution of the $\twolr$ chiral lagrangian is
\eqn\skyrmesoln{
\Sigma = A\ \Sigma_0\ A^{-1} ,
}
where
\eqn\skyrmezero{
\Sigma_0 = \exp\left[ i F(\abs{\vec x})\ \hat x \cdot \vec \tau\right],
}
and $A\in SU(2)$ is the collective coordinate associated with isospin
transformations of the soliton solution $\Sigma_0$. The radial shape function
$F(r)$ satisfies $F(0)=-\pi$ and $F(\infty)=0$ for a soliton with baryon number
one.\foot{The baryon number current
in Ref.~\anw\ should have the opposite sign, which changes
the sign of $F(0)$ for a baryon relative to that used in \anw.}
The soliton shape $F(r)$ depends on the details of higher derivative terms
in the chiral lagrangian. In the quantum theory, baryons have wavefunctions
that are functions of the matrix $A\in SU(2)$. The wavefunctions are
\eqn\wavefunction{
\psi_{R a m}(A)=(-1)^{R+m}\sqrt {\dim R}\ D^{*(R)}_{a\ -m}(A),
}
for a state with isospin $I=R$, $I_3=a$, $J=R$, and $J_3=m$. The matrices
$D^{(R)}$ are the representation matrices for $SU(2)$, and we have normalised
the measure on the $SU(2)$ group so that
\eqn\measure{
\int_{SU(2)} dA =1.
}
In QCD, the only soliton states in the theory are those with $2I=2J={\rm odd}$.

In the large $N_c$ limit, the soliton is very heavy, and the semiclassical
approximation is valid. In this limit, time derivatives can be neglected to
leading order, so that the interaction hamiltonian is
\eqn\intham{
H_I = - {ig\over 2} \int d^3 \vec x \
\Tr \bar H H \gamma^j \gamma_5
[\Sigma^{\dagger} \partial_j \Sigma] + \ldots\,\, ,
}
with $\Sigma$ given by Eq.~\skyrmesoln. In the limit that the $H$ field is very
heavy, the interaction energy is determined by the value of Eq.~\intham\ with
the $H$ field at the origin \jmw. Using the expansion
\eqn\fexp{
F(r) = F(0) + r F'(0) + \ldots =  -\pi + r F'(0) + \ldots,
}
in Eq.~\skyrmezero\ gives
\eqn\sexp{
\Sigma_0= - 1 - i \vec \tau \cdot \vec x\ F'(0)+\ldots,
}
so that the interaction hamiltonian is
\eqn\hintexp{\eqalign{
H_I &= {g F'(0)\over 2} \int d^3 \vec x \
\Tr \bar H H \gamma^j \gamma_5
A \tau^j A^{-1}\cr &=
{g F'(0)\over 4} \int d^3 \vec x \
\Tr \bar H H\gamma^j \gamma_5 \tau^k
\ \Tr A \tau^j A^{-1} \tau^k.
}}
The isospin operator on the $H$ field is
\eqn\hiso{
I^k_H\ H = - H {\tau^k\over 2},
}
and the spin operator of the light degrees of freedom acting on $H$ is
\eqn\hlightspin{
S^k_{\ell H}\ H = - H {\sigma^k\over 2}.
}
Thus the interaction hamiltonian of Eq.~\hintexp\ can be written in the form
\eqn\nint{
H_I =
{g F'(0)}\ I^k_H\ S^j_{\ell H}\
\Tr A \tau^j A^{-1} \tau^k,
}
using the fact that $H\gamma^j\gamma^5=-H\sigma^j$ in the rest frame of the $H$
field, and noting that $-\Tr \bar H H$ creates $H$ particles with probability
$+1$. The interaction hamiltonian gives a binding energy which is of order
$N_c^0$. The total energy of the bound state is the interaction energy plus the
mass of soliton (which is of order $N_c$) and the mass of the $D$ meson (which
is of order $N_c^0$).
Note that the interaction hamiltonian also produces a distortion in the shape
function $F$ in the presence of an $H$ particle.\foot{This was also
noted by J. Hughes.} However, this is an effect of order $1/N_c$,
since the coupling constant $g$ is of order one, whereas the terms in
the lagrangian with no $H$ field are of order $N_c$.

The problem of determining the energy of heavy-meson--soliton bound states is
reduced to the problem of determining the spectrum of the interaction
hamiltonian Eq.~\nint. The spectrum will respect the heavy quark symmetry,
since $\vec S_Q$ commutes with $H_I$. In Ref.~\jmw, the soliton states were
restricted to the nucleon subspace. In this case, the operator
$\Tr A \tau^j A^{-1} \tau^k$ can be replaced by $-8\, I_\Sigma^k \
S_{\ell\Sigma}^j/3$, where $I_\Sigma$ and $S_{\ell\Sigma}$ are the isospin and
spin of the light degrees of freedom
operators acting on the soliton state \anw.
The spin operator in the soliton sector is the same as the operator for the
spin of the light degrees of freedom in the soliton sector, since the soliton
does not contain any heavy quarks. Thus, in this case we find that
\eqn\intnuc{\eqalign{
H_I &=
-{ 8 g F'(0)\over 3}\ I^k_H\ S^j_{\ell H}\ I^k_\Sigma\ S^j_{\ell \Sigma} \cr
&=-{2 g F'(0)\over 3} \left(I^2 - I^2_H - I^2_\Sigma\right)
\left( S_\ell^2 - S^2_{\ell H} - S^2_{\ell \Sigma}\right)\cr
&=-{2 g F'(0)\over 3} \left(I^2 - 3/4\right)
\left( S_\ell^2 - 3/4\right),
}}
using $I= I_\Sigma+I_H$, and $S_{\ell}= S_{\ell \Sigma}+S_{\ell H}$. This is
the result given in \jmw.

In the large $N_c$ limit, all the collective excitations of the soliton are
degenerate, and one cannot restrict the soliton states to the nucleon subspace.
The interaction hamiltonian will mix different isospin excitations of the
soliton. Since the interaction hamiltonian Eq.~\nint\ involves only the light
degrees of freedom of the $H$ field, it is convenient to treat the $H$ field as
having $s_\ell=1/2$, and combine the spin of the heavy quark with the spin of
the light degrees of freedom at the end of the calculation.
The interaction hamiltonian commutes with total isospin and total spin, so it
is useful to combine soliton states with the light degrees of freedom in $H$
into states which are eigenstates of $I$ and $S_\ell$,
\eqn\states{
\ket{I\, a\ s_\ell\, m ; R\ I_{H}\ s_{\ell H}}
= \sket{R\, b\, n}\ \hket{I_H\, c\ s_{\ell H}\, p}
\clebsch{R}{I_H}{I}{b}{c}{a}
\clebsch{R}{s_{\ell H}}{s_\ell}{n}{p}{m} ,
}
where $\sket{R\, b\, n}$ denotes a soliton with $I=s_\ell=R$, $I_3=b$, $s_{\ell
3}=n$, and $\hket{I_H\, c\ S_{\ell H}\, p}$ denotes the light degrees of
freedom of $H$ with
$I=I_H$, $I_3=c$, $S_{\ell}=s_{\ell H}$ and $s_{\ell 3}=p$. We only need to
consider the case where $I_H=s_{\ell H}=1/2$.
The soliton wavefunctions for the states $\sket{R\, a\, m}$ are given
explicitly in Eq.~\wavefunction. Throughout this paper, soliton states will be
denoted by $\sket{\ }$, the light degrees of freedom of the heavy meson field
$H$ will be denoted by $\hket{\ }$, and the bound state of the two will be
denoted by $\ket{\ }$.

We can now compute the matrix elements of $\nint$ between the states given in
Eq.~\states,
\eqn\intmatrix{\eqalign{
&\bra{I'\, a'\ s_\ell'\, m' ; R'\ I_{H}\ s_{\ell H}}  H_I
\ket{I\, a\ s_\ell\, m ; R\ I_{H}\ s_{\ell H}}
= {g F'(0)}\cr
\noalign{\smallskip}
&\qquad\times\clebsch{R}{I_H}{I}{b}{c}{a}
\clebsch{R}{s_{\ell H}}{s_\ell}{n}{p}{m}
\clebsch{R'}{I_H}{I'}{b'}{c'}{a'}
\clebsch{R'}{s_{\ell H}}{s_\ell'}{n'}{p'}{m'}\cr
\noalign{\medskip}
&\qquad\times\sbra{R'\, b'\, n'}\Tr A \tau^j A^{-1} \tau^k
\sket{R\, b\, n}\ \hbra{I_H\, c'\ s_{\ell H}\, p'}\ I^k_H\ S^j_{\ell H} \
\hket{I_H\,
c\ s_{\ell H}\, p}.
}}
The matrix elements in the $H$ sector are the matrix elements of the generators
of $SU(2)$ in the given irreducible representation,
\eqn\mhsec{
\hbra{I_H\, c'\ s_{\ell H}\, p'}\ I^k_H\  S^j_{\ell H}\ \hket{I_H\, c\ s_{\ell
H}\, p}=
T^{k(I_H)}_{c' c}\ T^{j(s_{\ell H})}_{p' p},
}
where $T^{k(R)}$ is the generator in the irreducible representation $R$.
The generators can be written in terms of Clebsch-Gordan coefficients using the
Wigner-Eckart theorem,
\eqn\genclebsch{
T^{k(R)}_{ba}=\sqrt{R (R+1)}\clebsch R1Rakb.
}
The matrix elements in the soliton sector can be evaluated in terms of the
representation matrices of the adjoint representation
using the identity
\eqn\triden{
\Tr A \tau^j A^{-1} \tau^k = 2\ D^{(1)}_{k j}(A).
}
The product of two representation matrices can be written in terms of a single
representation matrix,
\eqn\dproduct{
D^{(R)}_{ab}(A)\ D^{(S)}_{cd}(A) = \clebsch RSTace \clebsch RSTbdf
D^{(T)}_{ef}(A),
}
and the integral over the $SU(2)$ group of the product of three $D$ matrices
can be evaluated by using Eq.~\dproduct\ and the orthogonality relation
\eqn\dorth{
\int_{SU(2)} D^{(R)}_{ab}(A)\ D^{(S)*}_{cd}(A) = {1\over\dim R}\ \delta_{RS}
\delta_{ac}\delta_{bd}.
}
The soliton matrix element needed for Eq.~\intmatrix\ is then
\eqn\mssec{\eqalign{
\sbra{R'\, b'\, n'}&\Tr A \tau^j A^{-1} \tau^k
\sket{R\, b\, n}
=2\ (-1)^{R+n+R'+n'}\sqrt{\dim R\ \dim R'}\cr
&\qquad\qquad\qquad\qquad
\times\int_{SU(2)}D^{(R')}_{b'-n'}(A)\ D^{*(R)}_{b-n}(A)\ D^{(1)}_{k j}(A)\cr
\noalign{\smallskip}
&\qquad= 2\, \sqrt{{\dim R\over \dim R'}}
\clebsch R1{R'}bk{b'}\clebsch R1{R'}{-n}j{-n'},
}}
using the fact that the adjoint representation is real.
The matrix element of the interaction hamiltonian has thus been evaluated in
terms of sums of products of eight Clebsch-Gordan coefficients by combining
Eqs.~\intmatrix--\mssec. These can be evaluated explicitly in terms of
$6j$-symbols to give
\eqn\hsixj{\eqalign{
&\bra{I'\, a'\ s_\ell' \, m' ; R'\ I_H \ s_{\ell H} } H_I
\ket{I\, a\ s_\ell\, m ; R\ I_H \ s_{\ell H} }
= \cr
&\qquad\qquad{- 2 g F'(0)}(-1)^{I+I_H+s_\ell+s_{\ell H}+2 R}\cr
&\qquad\times\sqrt{\dim R\,\dim R'\, \dim I_H\, \dim s_{\ell H}\, I_H(I_H+1)
s_{\ell H} (s_{\ell H}+1)}\cr
\noalign{\smallskip}
&\qquad\times\left\{\matrix{R&1&R'\cr I_H&I&I_H\cr}\right\}
\left\{\matrix{R&1&R'\cr s_{\ell H}&s_\ell&s_{\ell H}\cr}\right\}
\delta_{II'}\delta_{aa'}
\delta_{s_\ell s_\ell'}\delta_{mm'}.
}}
The interaction hamiltonian is diagonal in isospin and in the spin of the light
degrees of freedom.

It is now straightforward to determine the energies of the lightest heavy quark
baryon states. The light degrees of freedom have the spin and flavour quantum
numbers of an antiquark ($I_H=1/2$, $s_{\ell H}=1/2$), and will be denoted by
$\bar q$. Since $I_H$ and $s_{\ell H}$ in Eq.~\hsixj\ are both fixed to be
$1/2$ and the hamiltonian is diagonal in $a$ and $m$, we will drop those labels
from now on, and denote the states $\ket{I\, a\ s_\ell\, m ; R\ I_{H}=1/2\
s_{\ell H}=1/2}$ by
$\ket{I\ s_\ell ; R}$.
States which are linear combinations of $\ket{I\ s_\ell ;
R}$ for different values of $R$ will be denoted by $\kket{I\ s_\ell}$.
The nucleon $N$ has $I=1/2$, $s_\ell=1/2$ so the
$N\otimes \bar q$ states have $I=0,1$ and $s_\ell=0,1$. Similarly, the
$\Delta\otimes\bar q$ states have $I=1,2$ and $s_\ell=1,2$, \etc\ The
interaction hamiltonian mixes the $I=1,s_\ell=1$ states in the nucleon and
delta sectors, but the other states in the $N\bar q$ sector do not mix with any
other soliton-$\bar q$ states. The energies of the $\ket{0\ 0; \half}$,
$\ket{1\ 0; \half}$, and $\ket{0\ 1; \half}$ states are obtained from $\hsixj$
as $-3 g F'(0)/2$, $ g F'(0)/2$ and $g F'(0)/2$ respectively, which are the
binding energies given in Ref.~\jmw. In the $I=1$, $s_\ell=1$ channel, we have
the interaction hamiltonian in the
$\ket{1\ 1; \half}$ $\ket{1\ 1; \frac32}$ basis
\eqn\twostate{
H_I = -{ g F'(0)\over 6} \pmatrix{1&4\sqrt2\cr4\sqrt2&5\cr}.
}
In Ref.~\jmw, the $\Delta$ states were omitted, so that the energy of the
$\ket{1\ 1}$ state was the 11 matrix element in Eq.~\twostate,
$-g F'(0)/6$. In the large $N_c$ limit, where the $N$ and $\Delta$ are
degenerate, we have two $\ket{1\ 1}$ states which are the eigenstates of $H_I$
in Eq.~\twostate,
\eqn\twostates{\eqalign{
\kket{1\ 1}_0 &= \sqrt{\frac13}\ \ket{1\ 1; \half} + \sqrt{\frac23}\ \ket{1\ 1;
\frac32},\cr
\kket{1\ 1}_1 &= \sqrt{\frac23}\ \ket{1\ 1; \half} - \sqrt{\frac13}\ \ket{1\ 1;
\frac32},
}}
with energies $-3 g F'(0)/2$ and $g F'(0)/2$ respectively. The choice of
subscripts for the $\kket{1\ 1}$ states will be explained in Sec.~5.
In the large $N_c$ limit, we see that the states $\ket{0\ 0; \half}$ and
$\kket{1\ 1}_0$ are degenerate. The state $\ket{0\ 0; \half}$ when combined
with the heavy quark is the spin-1/2 $\Lambda_c$ baryon, and the state
$\kket{1\ 1}_0$ when combined with the the heavy quark is the degenerate
multiplet of the spin-1/2 $\Sigma_c$ and the spin-3/2 $\Sigma_c^*$. We see that
in the large $N_c$ limit, the $\Lambda_c$ and $\Sigma_c$ are degenerate. The
$\Sigma_c-\Lambda_c$ mass splitting is studied in the next section. We have
also found that the $\ket{0\ 1; \half}$, $\ket{1\ 0; \half}$ and $\kket{1\
1}_1$ states are degenerate.  These are exotic baryons with no quark
model analogues;  it is reassuring to see that, for positive $g$, they
are unbound.
One can use Eq.~\hsixj\ to compute the interaction
hamiltonian in the different $I,\ s_\ell$ sectors. All states in the spectrum
have an energy of either $-3 g F'(0)/2$ or $g F'(0)/2$.
The reason for this degeneracy will be explained in Sec.~5.

\newsec{The $\Sigma_c-\Lambda_c$ Mass Splitting}

The $\Sigma_c$ and $\Lambda_c$ are degenerate at leading order in $1/N_c$, but
at subleading order there are two terms which break the degeneracy.
The first
is the rotational kinetic energy of the soliton which splits the
nucleon-delta degeneracy. This term has a coefficient of order $N_c$ in the
lagrangian, and has two time derivatives. Since each time derivative brings a
factor of $1/N_c$ suppression, this term produces an energy splitting of order
$1/N_c$. The degeneracy is also broken by the second term in
Eq.~\intlag. Its coefficient is of order $N_c^0$ and it has one time
derivative, so it too produces an energy splitting of order $1/N_c$.
However, for the $SU(2)$ soliton
$$
\Sigma^\dagger {d\over dt}\Sigma
$$
vanishes at the origin where the $H$ particle is bound. Thus to
leading order in the derivative expansion the interaction term can be
neglected, and only the $N-\Delta$ splitting contributes.  The interaction
hamiltonian in the $\ket{1\ 1; \half},\ket{1\ 1; \frac32}$ sector can be
written as
\eqn\twomod{
H_I = -{ g F'(0)\over 6} \pmatrix{1&4\sqrt2\cr4\sqrt2&5\cr}
+\pmatrix{0&0\cr0&\Delta M\cr},
}
where $\Delta M$ is the $\Delta-N$ mass difference.

We will determine the
energies of the lowest heavy quark baryons using the experimental value for the
$\Delta-N$ mass difference and adjusting the unknown value of $g F'(0)$ to fit
the observed value of the $\Lambda_c$ mass,
\eqn\lcmass{
m_{\Lambda_c} = 2285\ {\rm MeV} = m_N + m_H - 3 g F'(0)/2,
}
which implies $g F'(0)=419$~MeV using the weighted average for the $D-D^*$
multiplet of $m_H=(3 D^*+D)/4=1975$~MeV. Thus we find that the energies of the
$\ket{1\ 0; \half}$ and $\ket{0\ 1; \half}$ states are 3124~MeV. These states
are exotic states, and the soliton model predicts that they are unbound.
The eigenvalues of \twomod\ are, to first order in $\Delta M$,
\eqn\evalues{\eqalign{
&E_-=-{3 g F'(0)\over 2} + {2\over 3} \Delta M\cr
&E_+={g F'(0)\over 2} + {1\over 3} \Delta M.}}
Thus one of the $\kket{1\ 1}$ states is unbound, and we have the
relation\foot{This result is true in the
constituent quark model, and was also obtained in the
Skyrme model in Ref.~\calkleb. However, Callan and Klebanov
found a $\Sigma_c^*-\Sigma_c$ mass difference in the heavy quark
limit because they did not include $D^*$ mesons.}
\eqn\callkleb{\bar m_{\Sigma_c}-m_{\Lambda_c}={2\over 3}\Delta M,}
where we have defined
$\bar m_{\Sigma_c}$ as the $\Sigma_c-\Sigma_c^*$ multiplet
average mass $(2m_{\Sigma_c^*}+m_{\Sigma_c})/3$.
This gives the
parameter free prediction
\eqn\paramfree{\bar m_{\Sigma_c}=2480\ \MeV.}
The $\Sigma_c^*$ mass has
not been measured, but the experimental value of the $\Sigma_c$ mass is
$2453$~MeV, so this agrees well with experiment.

We have computed the energies of baryons containing a heavy quark by expanding
the effective action in derivatives. In the semiclassical expansion, time
derivatives are considered to be small, but space derivatives are not
necessarily small. Terms in the interaction hamiltonian with no time
derivatives but an arbitrary number of space derivatives can produce only two
kinds of interaction terms, which can be written in the form
\eqn\terms{
H_I=V_0(F)\, +\, V_1(F)\ \Tr A\tau^j A^{-1}\tau^k\ I^k_H\ S^j_{\ell H},
}
where $V_0$ and $V_1$ are functionals of the soliton shape function $F$ and are
of order $N_c^0$. The leading contribution to $V_1$ is first order in
derivatives, $V_1=g F'(0)$, as we determined in Eq.~\nint. The leading
contribution to $V_0$
is second order in space derivatives, \eg\ from a term of the form
$\Tr \bar H H\ \Tr \partial^\mu \Sigma \partial_\mu \Sigma^\dagger$.
Terms with one time derivative and two $H$ fields are of order $1/N_c$, and can
be written in the form
\eqn\onetime{
H_I = V_2(F)\ I^k_H\ I^k_\Sigma,
}
where $V_2$ is of order $1/N_c$.
The leading term with one time derivative and no space derivatives from
Eq.~\intlag\ vanishes, so $V_2(F)$ starts at second order in space derivatives.
Terms with two time derivatives and no $H$ fields have the form
$$
\lambda(F)\ I^2_\Sigma\ ,
$$
where the functional $1/2\lambda$ is the moment of inertia of the soliton,
starts off at zero space derivatives, and is of order $N_c$. The value of
$\lambda$ can be chosen to fit the observed $N-\Delta$ mass difference. That
leaves three parameters $V_0$, $V_1$ and $V_2$ to fit the observed $\Lambda_c$
and $\Sigma_c$ masses, so there is no predictive power. If we assume that we
also have a systematic expansion in space derivatives, then we can neglect
$V_0$ and $V_2$ relative to $V_1$, and we get the predictions discussed in the
preceding paragraph. Finally, one can use the value $V_1=gF'(0)$ combined with
the slope of the shape function $F'(0)\simeq 700$~MeV used in Ref.~\anw\ to get
$g^2\simeq 0.36$ compared with the experimental limit $g^2 < 0.4$.

\newsec{Masses in the Large $N_c$ Limit: The Quantum Number $K$}

In the previous section, we noticed that the $\ket{0\ 0; \half}$ and $\kket{1\
1}_0$ states were degenerate, as were the $\ket{1\ 0; \half}$, $\ket{0\ 1;
\half}$ and $\kket{1\ 1}_1$ states. We will derive this result by a different
method, which is useful in generalising the results to $SU(3)$. The interaction
hamiltonian we consider has the form
\eqn\newhint{
H_I = V_1(F)\ I^k_H\ S^j_{\ell H}
\ \Tr A \tau^j A^{-1} \tau^k,
}
with $V_1$ an arbitrary functional of $F$.
In the previous section, we diagonalised this hamiltonian by considering states
$\ket{I\  s_\ell  ; R}$ which were eigenstates of total isospin and total
light-spin built from products of soliton and $\bar q$ states. In this section,
we consider instead the states
\eqn\newstates{
\sket{\Sigma_0}\hket{a\, m}
}
which are the tensor product of a soliton with $A=1$\foot{\ie\ the wave
function of the collective coordinate is $\psi(A)=\delta(A=1)$}\
and a $\bar q$ state with
$I_3=a$ and $s_{\ell 3}=m$. Acting on this state with $H_I$ gives
\eqn\honstate{
H_I \sket{\Sigma_0}\hket{a\, m}= V_1\,(2\ I_H \cdot S_{\ell
H})\sket{\Sigma_0}\hket{a\, m}.
}
It is convenient to define $\vec K=\vec I + \vec S_{\ell}$, and use $\bar q$
states which are representations of $K$.\foot{Many of the Skyrme model
predictions follow from invariance of $\sket{\Sigma_0}$ under $K$. This
invariance was used in Ref.~\am\ to show that
the quark model and Skyrme model are
equivalent in the large $N_c$ limit. It has also been used to
obtain relations between
pion-nucleon scattering amplitudes
\ref\mattis{M.~Mattis and M.~Mukerjee, \prl{61}{1988}{1344}}. $K$ is
equal to $I+S_\ell$ only for solitons $\sket{\Sigma_0}$ which have
$A=1$. For the generic soliton configuration $\sket{A\Sigma_0 A^{-1}}$
with collective coordinate $A$, $K$ is equal to $\CR(A)\,I\,\CR(A^{-1}) +
S_\ell$ where $\CR(A)$ is an isospin rotation by $A$. Terms in the
chiral lagrangian with no time derivative are invariant under $K$.}
Since $\bar q$ has $I_H=1/2$ and $s_{\ell H}=1/2$, we have states with $K=0,1$,
which will be labeled $\hket{K\, k}$. This allows us to diagonalise the
interaction hamiltonian,
\eqn\hk{
H_I \sket{\Sigma_0}\hket{K\, k}= V_1\, (2\, I_H \cdot S_{\ell
H})\sket{\Sigma_0}\hket{K\, k} = V_1\,
(K_H^2 - I_H^2 - S_{\ell H}^2) \sket{\Sigma_0}\hket{K\,
k}.
}
We see that the states with $K=1$ have energy $V_1/2$ and that states with
$K=0$ have energy $-3V_1/2$, which are precisely the allowed energies found in
the previous section. We can then find states with definite values of total
isospin and light-spin by applying isospin and light-spin projection operators
to $\sket{\Sigma_0}\hket{K\, k}$ \am,
\eqn\eigen{
\ket{I\, a\ s_\ell\, m; K\, k\, b\, m}
\propto P^R_{ab}(I) \ P^{s_\ell}_{mn}\
(S_\ell)\ \sket{\Sigma_0}\ \hket{K\, k}
}
where the $SU(2)$ projection operators are defined by
\eqn\proj{
P^R_{ab}(X) = \int_{SU(2)} dg\ D^{*(R)}_{ab}(g)\ \hat U_X(g),
}
where $\hat U_X(g)$ is the group transformation operator with generator $X$.
The isospin projector $\hat U_I$ is the exponential of the isospin generators,
and the spin projector $\hat U_{S_\ell}$ is the exponential of the spin
generators of the light degrees of freedom. We will also need $\hat U_K$ which
is the exponential of $K=I+S_\ell$.
The analysis of the allowed states is now a generalisation of the results of
\am. The soliton $\sket{\Sigma_0}$ is invariant under the action of
$I+S_{\ell}$, which places constraints on which states can be projected out in
Eq.~\eigen\ because of the identity
\eqn\useful{
\hat U_I(g)\ \hat U_{S_\ell} (h) \sket{\Sigma_0} = \hat
U_I(gh^{-1})\sket{\Sigma_0}.
}
The isospin operator can be written as the product of the isospin operators on
the soliton and $\bar q$, and similarly for the spin operators, so that
\eqn\moreproj{\eqalign{
\ket{I\, a\ s_\ell\, m; K\, k\, b\, n} &=
\int_{SU(2)} dg\ D^{*(R)}_{ab}(g)
\int_{SU(2)} dh\ D^{*(s_\ell)}_{mn}(h)\cr
\noalign{\smallskip}
&\qquad\times\hat U_I(g)\ \hat U_{S_\ell}(h)\sket{\Sigma_0}\ \hat U_I(g)\ \hat
U_{S_\ell}(h)
\hket{K\, k}\cr
\noalign{\medskip}
&=\int_{SU(2)} dg\ D^{*(R)}_{ab}(g)
\int_{SU(2)} dh\ D^{*(s_\ell)}_{mn}(h)\cr
\noalign{\smallskip}
&\qquad\times\hat U_I(gh^{-1})\sket{\Sigma_0}\ \hat U_I(g)\ \hat U_{S_\ell}(h)
\hket{K\, k},\cr
}}
using Eq.~\useful. Replacing the dummy group element $g$ by $gh$, and using the
group property of the representation matrices and the $\hat U$ operators, we
get
\eqn\projtwo{\eqalign{
\ket{I\, a\ s_\ell\, m; K\, k\, b\, n} &=
\int_{SU(2)} dg \int_{SU(2)} dh\
D^{*(R)}_{ac}(g)\  D^{*(R)}_{cb}(h)\
D^{*(s_\ell)}_{mn}(h)\cr
\noalign{\smallskip}
&\qquad\times\hat U_I(g)\sket{\Sigma_0}\ \hat U_I(g)\ \hat U_I(h) \hat
U_{S_\ell}(h) \hket{K\, k}.
}}
We also have the identity
$$
U_I(h)\, U_{S_\ell} (h)\ \hket{K\, k} = U_K(h)\ \hket{K\, k} = \hket{K\, k'}
D^{(K)}_{k'k}(h),
$$
since $K=I+S_\ell$, and $\hket{K\, k}$ is an irreducible representation under
$K$.
Thus Eq.~\projtwo\ can be simplified to
\eqn\projthree{\eqalign{
\ket{I\, a\ s\, m; K\, k\, b\, n} &=
\int_{SU(2)} dg \int_{SU(2)} dh\
D^{*(R)}_{ac}(g)\  D^{*(R)}_{cb}(h)\ D^{(K)}_{k'k}(h)\
D^{*(s_\ell)}_{mn}(h)\cr
\noalign{\smallskip}
&\qquad\qquad\qquad\times\hat U_I(g)\sket{\Sigma_0}\ \hat U_I(g) \hket{K\,
k'}\cr
\noalign{\medskip}
&={1\over \dim K}\clebsch I{s_\ell}Kcm{k'}\clebsch I{s_\ell}Kbnk\int_{SU(2)} dg
\ D^{*(R)}_{ac}(g)\cr
\noalign{\smallskip}
&\qquad\qquad\qquad\times\hat U_I(g)\sket{\Sigma_0}\hat U_I(g) \hket{K\,
k'}.\cr
}}
Multiplying both sides of the equation by $\clebsch I{s_\ell}Kbnk$, we find
\eqn\sfinal{\eqalign{
\kket{I\, a\ s_\ell\, m}_K&=\lambda\clebsch I{s_\ell}Kbnk\ket{I\, a\ s_\ell\,
m; K\, k\, b\, n}
\cr&=\lambda
\int_{SU(2)} dg\
D^{*(I)}_{ac}(g)\
\hat U_I(g)\sket{\Sigma_0}\hat U_I(g) \hket{K\, k}
\clebsch I{s_\ell}Kcm{k},
}}
which defines the state $\kket{I\, a\ s_\ell\, m }_K$. $\lambda$ is a
normalisation constant which has been inserted so that the state has unit norm.
To determine $\lambda$, consider the overlap
\eqn\sover{\eqalign{
&\phantom{\kket{}}_{K'}\!\bbra{I'\, a'\ s_\ell'\, m'}\!\kket{I\, a\ s_\ell\,
m}_K =
\lambda'\lambda\int_{SU(2)} dg\ \int_{SU(2)} dg'\
D^{*(I)}_{ac}(g)\  D^{*(I')}_{a'c'}(g')\cr
\noalign{\smallskip}
&\qquad\qquad\times\sbra{\Sigma_0}\hat U_I(g')^{-1}
\hat U_I(g)\sket{\Sigma_0}\
\hbra{K'\, k'}\hat U_I(g')^{-1} \hat U_I(g) \hket{K\, k}\cr
\noalign{\smallskip}
&\qquad\qquad\qquad\qquad\times\clebsch
I{s_\ell}Kcm{k}\clebsch{I'}{{s_\ell}'}{K'}{c'}{m'}{k'}.
}}
The orthogonality of soliton states with different rotations reduces the double
integral to a single integral,
\eqn\soverii{\eqalign{
&\phantom{\kket{}}_{K'}\!\bbra{I'\, a'\ {s_\ell}'\, m'}\!\kket{I\, a\
{s_\ell}\, m}_K =
\lambda'\lambda\int_{SU(2)} dg\
D^{*(I)}_{ac}(g)\  D^{*(I')}_{a'c'}(g)\cr
\noalign{\smallskip}
&\qquad\qquad\qquad\times\clebsch
I{s_\ell}Kcm{k}\clebsch{I'}{{s_\ell}'}{K'}{c'}{m'}{k'} \delta_{K K'}\delta_{k
k'}\cr
\noalign{\medskip}
&={\lambda'\lambda\over \dim I}\ \delta_{II'}\delta_{aa'}\delta_{cc'}
\delta_{K K'}\delta_{k k'}
\clebsch I{s_\ell}Kcm{k}\clebsch{I'}{{s_\ell}'}{K'}{c'}{m'}{k'}\cr
\noalign{\medskip}
&={\lambda'\lambda\ \dim K\over \dim I\ \dim s_\ell}\delta_{II'}\delta_{aa'}
\delta_{{s_\ell}{s_\ell}'}\delta_{mm'}\delta_{K K'},
}}
so that the normalisation factor $\lambda$ for $\kket{I\ {s_\ell}}_K$ is
\eqn\ijknorm{
\lambda = \sqrt {\dim I\ \dim s_\ell \over \dim K}.
}

The only states that one can construct are those for which the Clebsch-Gordan
coefficient in Eq.~\sfinal\ does not vanish, $K\subset I\otimes {s_\ell}$, and
have $2I={\rm even}$. The fermionic nature of $\bar q$ changes the
quantisation condition from $2I={\rm odd}$ for the soliton to $2I={\rm even}$
for the soliton-$\bar q$ composite state.
The subscript on the states $\kket{1\ 1}$ in Eq.~\twostates\ is the value of
$K$. For $K=0$, we have an infinite tower of states with
$I={s_\ell}=0,1,2,\ldots$ with energy $-3 V_1/2$. For $K=1$, we have three
degenerate infinite towers, $I={s_\ell}+1=1,2,3,\ldots$,
${s_\ell}=I+1=1,2,3,\ldots$, and $I={s_\ell}=1,2,3\ldots$ with energy $g
V_1/2$. The states $\ket{1\ 0;\half}$ and $\ket{0\ 1;\half}$ and $\kket{1\
1}_1$ of the previous section are from the $K=1$ series, and the states
$\ket{0\ 0;\half}$ and $\kket{1\ 1}_0$ are from the $K=0$ series.
The usual quantisation condition for the soliton (\ie\ without any heavy
mesons) is a special case of the above construction. We can recover the results
by omitting the $\bar q$ quantum numbers, so that $K=0$, and $2I = {\rm odd}$.
This gives the usual tower $I={s_\ell}=1/2,3/2,5/2,\ldots$.

\newsec{Pion-Baryon Coupling Constants}

The soliton model can also be used to determine the pion coupling constants of
the baryons in terms of the pion coupling constant of the heavy meson, $g$.
These can be evaluated using the matrix element of the axial vector current in
the soliton heavy meson bound state. We will evaluate the matrix element of the
axial current $j^{\mu A}$ at zero momentum transfer, with $\mu=3$ and $A=+$.
There are two independent coupling constants that determine all the pion
couplings. We will follow the notation of Cho \ref\cho{P. Cho, Harvard Preprint
HUTP-92/A014 (1992)}\ who denoted the couplings by $g_2$ and $g_3$. ($g_1$ as
defined by Cho is equal to the meson pion coupling constant $g$ used in this
paper.)   To determine $g_2$ and $g_3$ we need to compute two
independent matrix
elements of $j^{3+}$.
This axial current can be written in terms of baryon fields as
\eqn\baxial{
j^{3+}=-{\sqrt 2\over 3} g_2\ \bar{\Sigma_c}^{++}\sigma_3\Sigma_c^+
- {1\over \sqrt3} g_3\ \bar{\Lambda_c}^+\sigma_3\Sigma_c^0 + \ldots.
}
where we have used the lagrangian given in Ref.~\cho\ and retained only the
terms we will use in the calculation.

The axial current in the chiral lagrangian is
\eqn\caxial{
j^{3+} = {\pi D\over 2 e^2} \Tr A \tau^3 A^{-1} \tau^+ - g
\Tr\bar H H \gamma^3\gamma_5 \left[{\Sigma^\dagger\tau^+\Sigma+\tau^+\over
2}\right] + \ldots,
}
where the ellipsis denotes higher derivative terms in the chiral lagrangian.
The first term in the axial current is from the purely pionic sector of the
chiral lagrangian, and is identical to that studied by Adkins, Nappi and Witten
\anw, and the constants $D$ and $e$ are defined in their paper.\foot{
We have
included the factor of 3/2 mentioned before Eq.~(20) of Ref.~\anw\ into the
definition of the axial current.
There is a correction to the expression for $D$ given in Ref.~\anw\
because of the distortion of the shape function $F$
in the presence of the $H$ field.  However, as mentioned in Sec.~3,
this is a $1/N_c$ effect proportional to $g$, and we will neglect it. The
distortion of $F$ also changes the long distance behaviour of
the soliton solution so
that the Goldberger-Treiman relation is satisfied for the bound state.
}
The second term is the axial current of the
$H$ field from  Eq.~\intlag. In the heavy quark limit, $\Sigma$ can be replaced
by its value at the origin, $-1$.

The two independent matrix elements that we will compute are
\eqn\tmelem{\eqalign{
\bra{\Sigma_c^{++}\uparrow}j^{3+}\ket{\Sigma_c^+\uparrow} &= -{\sqrt 2\over 3}
g_2,\cr
\bra{\Lambda_c^+\uparrow}j^{3+}\ket{\Sigma_c^0\uparrow} &=-{1\over \sqrt3} \
g_3, \cr
}}
using Eq.~\baxial. The same matrix elements computed using soliton heavy meson
bound states will determine $g_2$ and $g_3$. The states we have constructed so
far have omitted the spin of the heavy quark. To obtain the $\Lambda_c$ state,
we need to combine the heavy quark with the state $\kket{0\ 0}_0$, to obtain
\eqn\lstate{
\ket{\Lambda_c\uparrow} = \kket{0\ 0}_0\ket{\uparrow}_Q,
}
where $\ket{\ }_Q$ denotes the heavy quark spin state.
Similarly, the $\Sigma_c$
states are obtained by combining the heavy quark spin
with the state $\kket{1\ 1}_0$,
\eqn\sstate{
\ket{\Sigma_c \uparrow} = \sqrt{\frac23}\ \kket{1\ 1\, s_{\ell
3}=1}_0\ket{\downarrow}_Q - \sqrt{\frac13}\ \kket{1\ 1\, s_{\ell
3}=0}_0\ket{\uparrow}_Q .
}
The state $\Sigma_c$ has been chosen to be the state $\kket{1\ 1}_0$, which is
the leading contribution in the large $N_c$ limit.

The matrix element of the first term in Eq.~\caxial\ uses the identity
$$
A\ \hat U_I(g)\sket{\Sigma_0} = g\ \hat U_I(g)\sket{\Sigma_0},
$$
where $A$ is the soliton collective coordinate,
and Eq.~\triden. It is convenient to label the $\tau$ matrices by the angular
momentum indices $\pm 1,0$, instead of the cartesian labels $x,y,z$. The
relation between the two bases is
$\tau^{\pm 1}=\mp(\tau^x\pm i\tau^y)/\sqrt2$ and $\tau^0=\tau^3$. In the
angular momentum basis, Eq.~\triden\ becomes
$$
\Tr A \tau^b A^{-1} \tau^a = 2\ (-1)^b\, D^{*(1)}_{a-b}(A).
$$
The expression for the $\kket{\ }_K$ states, Eq.~\sfinal\ can be simplified for
the case $K=0$,
\eqn\ggai{\eqalign{
\kket{I\, a\ {s_\ell}\, m}_0&=\sqrt{\dim I}\ (-1)^{s_\ell+m}\int_{SU(2)} dg\
D^{*(I)}_{a-m}(g)\
\hat U_I(g)\sket{\Sigma_0}\ \hat U_I(g) \hket{K=0},
}}
where we must have $I={s_\ell}$ since $K=0$.
Using this simplified expression for the states gives the matrix element
\eqn\gai{\eqalign{
&\phantom{\kket{}}_{0}\!\left.\bbra{I'\,a'\ {s_\ell}'\,m'}\right|
\Tr A\tau^r A^{-1}\tau^s\kket{I\,a\ {s_\ell}\,m}_0=2\
(-1)^{r+s_\ell+m-s_\ell'-m'}
\cr
&\qquad\qquad\times
\sqrt {\dim I\ \dim I'}\int_{SU(2)} dg
D^{*(I)}_{a-m}(g)\ D^{*(1)}_{s-r}(g)\  D^{(I')}_{a'-m'}(g)\cr
\noalign{\smallskip}
&\qquad\qquad=(-2) \sqrt {\dim I\over \dim I'}
\clebsch I1{I'}as{a'}\clebsch{I}{1}{I'}{m}{r}{m'}.\cr
}}
Thus we obtain the matrix element of the first term in Eq.~\caxial,
\eqn\gaii{\eqalign{
\bra{\Sigma_c^{++}\uparrow}j^{3+}\ket{\Sigma_c^+\uparrow} &= -{\pi D\over
3\sqrt2 e^2}, \cr
\bra{\Lambda_c^+\uparrow}j^{3+}\ket{\Sigma_c^0\uparrow} &={\pi D\over 3\sqrt2
e^2}.
}}

The second part of the axial current can be written in the form
\eqn\gaiii{
-g \Tr\bar H H \gamma^r\gamma_5 \tau^s = - 2 g\ S^r_{\ell H}\ I^s_{H},
}
so that we need to evaluate the matrix element:
\eqn\gaiv{\eqalign{
\noalign{\smallskip}
&\phantom{\kket{}}_{0}\!\left.\bbra{I'\,a'\ {s_\ell}'\,m'}\right|
S^r_{\ell H} I^s_H\kket{I\,a\ {s_\ell}\,m}_0=
\sqrt {\dim I\ \dim I'}\ (-1)^{I+m-I'-m'}\cr
\noalign{\smallskip}
&\qquad\times\int_{SU(2)} dg\
D^{*(I)}_{a-m}(g)\  D^{(I')}_{a'-m'}(g)\
\hbra{K=0}\hat U_I(g^{-1}) S^r I^s \hat U_I(g)\hket{K=0}\cr
\noalign{\medskip}
&=
\sqrt {\dim I  \over \dim I'}\ (-1)^{I+m-I'-m'}
\hbra{K=0}S^r_{\ell H} I^p_H\hket{K=0}
\cr
\noalign{\smallskip}
&\qquad\qquad\qquad\qquad\times
\clebsch I1{I'}as{a'}\clebsch{I}{1}{I'}{-m}{p}{-m'},\cr
}}
since
$$
\hat U_I(g^{-1})\ S^r_{\ell H}\ I^s_H\ \hat U_I(g) = S^r\ I^p\
D^{*(1)}(g)_{sp}.
$$
The operators  $S^r$ and  $I^s$ are irreducible tensor operators under $K$ with
$K=1$, so their product can have $K=0,1,2$. Only the $K=0$ part of $S^r\ I^s$
has a non-zero matrix element between $K=0$ states, so we can make the
replacement
\eqn\gavi{\eqalign{
\hbra{K=0}S^r_{\ell H} I^p_H \hket{K=0}\rightarrow&{1\over3}(-1)^r\delta_{p-r}
\hbra{K=0}S_{\ell H}\cdot I_H \hket{K=0}\cr
=& -{1\over4}(-1)^r\delta_{p-r}
}}
since $2\ S_{\ell H}\cdot I_H = K_H^2 - I^2_H - S^2_{\ell H} = -3/2$.
This allows the sum in Eq.~\gaiv\ to be evaluated,
\eqn\gav{\eqalign{
&{}_{0}\!\left.\bbra{I'\,a'\ {s_\ell}'\,m'}\right|
S^r I^s\kket{I\,a\ {s_\ell}\,m}_0=
{1\over 4}\sqrt{\dim I\over \dim I'}\clebsch
I1{I'}as{a'}\clebsch{I}{1}{I'}{m}{r}{m'}.\cr
}}
The matrix element of $j^{3+}$ is obtained by setting $r=0$ and $s=+$ in
Eq.~\gav\ and multiplying by $(-2g)(-\sqrt 2)$, since $I^+=-\sqrt 2 I^1$. This
gives the matrix elements of the second term of Eq.~\caxial,
\eqn\ggavi{\eqalign{
\bra{\Sigma_c^{++}\uparrow}j^{3+}\ket{\Sigma_c^+\uparrow} &= -{g\over 3\sqrt2},
\cr
\bra{\Lambda_c^+\uparrow}j^{3+}\ket{\Sigma_c^0\uparrow} &={g\over 3\sqrt2}.
}}
Combining Eqs.~\gaii\ and \gavi, and comparing with Eq.~\baxial, we get
\eqn\gavii{\eqalign{
g_2 = {\pi D\over 2e^2} + {g\over 2}&=-{3\over2}g_A +{g\over 2},\cr
\noalign{\smallskip}
g_3 = -{\pi D\over \sqrt6 e^2} - {g\over \sqrt6}&= \sqrt{{3\over2}}g_A
-{g\over \sqrt6},
}}
since the axial coupling constant of the nucleon is
\eqn\nucax{
g_A = -{\pi D \over 3 e^2}.
}
Using $g_A=1.25$, and setting $g$ equal to its experimental upper bound
\accmor\ of $0.63$, we get
\eqn\gaix{
g_2 = -1.6,\quad g_3 = 1.3\ .
}

To all orders in the derivative expansion, the $g_A$ term has the form given in
Eq.~\caxial, where $D$ is some functional of the shape function $F$. The value
of $D$ is normalised using the formula for $g_A$ in Eq.~\nucax. The $g$ term
gets corrections from higher derivative terms in the effective action. These
higher derivative terms retain the flavour structure of the leading term
$\Tr\bar H H \sigma^3\tau^+$, but renormalise the coefficient so that it is no
longer $g$. This renormalisation of $g$ cancels in the ratio, so that we have
the large $N_c$ prediction
\eqn\gaviii{
{g_2\over g_3}= - \sqrt{{3\over 2}}.
}
In addition, since $g$ is formally of order $N_c^0$ whereas $g_A$ is of order
$N_c$, the leading $N_c$ predictions for the values of $g_2$ and $g_3$ are
given by retaining only the $g_A$ term in Eq.~\gavii,
which leads to the model independent
large $N_c$ prediction
\eqn\gaxx{
g_2 = -{3\over 2}g_A=-1.9,\quad g_3 = \sqrt{3\over2}g_A=1.5\ .
}
The correction to Eq.~\gaxx\ is of order $1/N_c$ relative to the leading term,
and the correction to Eq.~\gaviii\ is of order $1/N_c^2$ relative to the
leading term.
Using the value \gaix\ for $g_3$ along with
the measured baryon masses gives the
prediction
\eqn\bwidth{
%\eqalign{\Gamma(\Sigma_c^{*++}\rightarrow\Sigma_c^+\pi^+)&=
%{g_2^2\over 36\pi}{|\vec p_\pi|^3\over f^2}\simeq
%( 1.3\times 10^{-6} \MeV^{-2}) \ |\vec p_\pi|^3, \cr
\Gamma(\Sigma_c^{++}\rightarrow\Lambda_c^+\pi^+)=
{g_3^2\over 6\pi}{|\vec p_\pi|^3\over f^2}\simeq 3.7\,\MeV.}
%}
The couplings $g_2$ and $g_3$ have also been computed in the constituent
quark model, and the prediction for the decay width Eq.~\bwidth\ is
between the values found in Ref.~\yan\ for the naive quark model, and
for the chiral quark model \ref\chiralquark{A.V. Manohar and H. Georgi,
\np{234}{1984}{189}}.

\newsec{Baryons containing a Heavy Quark: The $SU(3)$ Case}

It is straightforward to generalise the results of the previous section to the
case of $SU(3)$. The soliton in $SU(3)$ has the form Eq.~\skyrmezero\ where the
isospin matrices form a subgroup of $SU(3)$. The soliton is still invariant
under $K=I+{s_\ell}$. In addition, the Wess-Zumino term requires that the
soliton have a definite hypercharge \guad\am,
\eqn\yconsti{
3 Y\sket{\Sigma_0} = N_c \sket{\Sigma_0},
}
where we are interested in the case where the number of colours $N_c=3$. The
hypercharge generator is
$$
Y=\pmatrix{1/3&0&0\cr0&1/3&0\cr0&0&-2/3}.
$$

The soliton-$\bar q$ bound states can be obtained using the methods of the
previous section. The $\bar q$ state transforms as a $\bar 3$ under $SU(3)$ and
as a doublet under the spin of the light degrees of freedom. Under the
$SU(2)\times U(1)$ subgroup, it decomposes into an isospin doublet with
hypercharge $-1$, and an isospin singlet, with hypercharge 2. Thus we have
$\bar q$ states with $\{K=0,1;\ 3Y=-1\}$, and $\{K=1/2;\  3Y=2\}$. The energies
are still given by Eq.~\hk, so that we have a tower of states with energy
$V_1/2$ from the $K=1$ states, energy $-3V_1/2$ from the $K=0$ states, and
energy zero from the $K=1/2$ states.
The bound states can be computed by a procedure similar to that used in Sec.~5.
The details are uninteresting, and the final result is that the allowed states
are given by
\eqn\ssfinal{\eqalign{
\kket{R\, a\ {s_\ell}\, m}_K=&\sqrt{{\dim R\ \dim s_{\ell}\over \dim K}}
\int_{SU(3)} dg\
D^{*(R)}_{ac}(g)\cr
&\qquad\times\hat U_I(g)\sket{\Sigma_0}\hat U_I(g) \hket{K\, k'}
\clebsch {I_c}{s_\ell}Kcm{k'},
}}
with the additional constraint that
\eqn\yconstii{
Y_c = Y_\Sigma + Y_K = 1 + Y_K,
}
where $I_c$ and $Y_c$ are the isospin and hypercharge of the $SU(3)$ state $c$
of the representation $R$, and $Y_K$ is the hypercharge of the $\bar q$ state
with isospin $K$. Thus the allowed $SU(3)$ representations for a spin
${s_\ell}$ state are those which contain an element with isospin $I$ and
hypercharge $Y$ such that
\eqn\possib{\eqalign{
&(a)\ I={s_\ell} ,\ 3Y=2,\cr
&(b)\ 1\subset I\otimes {s_\ell},\ 3Y =2,\cr
&(c)\ \half\subset I\otimes {s_\ell},\ 3Y=5,\cr
}}
so that the Clebsch-Gordan coefficient in Eq.~\ssfinal\ does not vanish, and
the hypercharge constraint is satisfied. Cases (a)--(c) are from the $\{K=0,\
3Y_K=-1\}$, $\{K=1,\ 3Y_K=-1\}$, and $\{K=1/2,\ 3Y_K=2\}$ states of $\bar q$
respectively. The usual soliton quantisation condition for $SU(3)$ (\ie\ with
no heavy meson fields) is obtained by omitting the $\bar q$ state, \ie\ using
$\{K=0,Y_K=0\}$,
and is that $R$ must contain a state with $I={s_\ell}$  and with $Y=1$.

To classify the allowed states, we need to find the decomposition of
irreducible tensors of $SU(3)$ into the $SU(2)\times U(1)$ subgroup generated
by isospin and hypercharge. An irreducible $SU(3)$ representation $(m,n)$ is a
traceless tensor which is completely symmetric in $m$ upper and $n$ lower
indices, and has dimension
\eqn\dimmn{
\dim (m,n)={(m+1)(n+1)(m+n+2)\over 2}
}
If $[m,n]$ denotes the (reducible) tensor which is completely symmetric in its
upper and lower indices, then the representations in $(m,n)$ are those in
$[m,n]\ominus[m-1,n-1]$, because the trace of $[m,n]$ is the tensor
$[m-1,n-1]$. It is much simpler to compute the decomposition of $[m,n]$ because
we do not have the constraint that the tensor has zero trace. The tensor
$[m,n]$ under the $SU(2)\times U(1)$ subgroup decomposes into the sum
\eqn\dmn{
\sum_{k=0}^m\sum_{r=0}^n\ [m-k,n-r]_{3Y=m-n-3k+3r},
}
where the term $[m-k,n-r]$ is obtained by setting $r$ upper and $k$ lower
indices equal to 3. The remaining $m-k$ upper and $n-r$ lower indices can have
values 1,2. The tensor $[m-k,n-r]$ is symmetric on $m-k$ upper and $n-r$ lower
indices, and is the reducible representation given by the tensor product of the
representations $2I=m-k$ and $2I=n-r$ of $SU(2)$.
By subtracting the decomposition of $[m-1,n-1]$ from $[m,n]$, one obtains the
decomposition
\eqn\sudecomp{\eqalign{
(m,n)\mathop{\longrightarrow}\limits^{SU(2)}&\sum_{r=0}^{n-1}\ \left\{
{\abs{m-r}\over 2},{\abs{m-r}\over 2}+1,\ldots,{m+r\over
2}\right\}_{Y=m+2n-3r}\cr
\oplus&\sum_{r=0}^{m-1}\ \left\{
{\abs{n-r}\over 2},{\abs{n-r}\over 2}+1,\ldots,{n+r\over
2}\right\}_{Y=-n-2m+3r}\cr
&\oplus\ \left\{
{\abs{m-n}\over 2},{\abs{m-n}\over 2}+1,\ldots,{m+n\over 2}\right\}_{Y=m-n}.\cr
}}

The soliton heavy meson bound states are now constructed in two different ways.
We first construct the states by tensoring the soliton states with the $\bar q$
states. These states are the generalisation of the $\ket{I\ {s_\ell};R}$ states
of Sec.~3. We then construct the bound states directly using Eq.~\ssfinal. The
states obtained by this method are the generalisation of the $\kket{I\
{s_\ell}}_K$ states of Sec.~5.
The soliton states have $I={s_\ell}$ and $Y=1$. The lowest representations are
${\bf 8}=(1,1)$ with ${s_\ell}=1/2$, ${\bf 10}=(3,0)$ with ${s_\ell}=3/2$,
${\bf 27}=(2,2)$ with ${s_\ell}=1/2,3/2$, ${\bf 35}=(4,1)$ with
${s_\ell}=3/2,5/2$, ${\bf \bar{10}}=(0,3)$ with ${s_\ell}=1/2$, \etc\
The tensor product of the soliton with $\bar q$ can be computed using the
formula
\eqn\qtensor{
(m,n)\otimes (0,1) = (m,n+1) \oplus (m+1,n-1) \oplus (m-1,n).
}
Thus the lowest soliton heavy meson bound states are: ${\bf 8}\otimes{\bf{\bar
3}}$ = $(1,1)\otimes (0,1)$ = $(1,2) \oplus (2,0) \oplus (0,1)$ = ${\bf{\bar
{15}}}\oplus {\bf 6}\oplus {\bf{\bar 3}}$, and have light-spin ${s_\ell}=0,1$;
${\bf 10}\otimes {\bf{\bar 3}}$ = $(3,0)\otimes (0,1)$ = $(3,1)\oplus (2,0)$ =
${\bf 24}\oplus {\bf 6}$ and have light-spin ${s_\ell}=1,2$, \etc\
The quark model states are the ${\bf\bar3}=(0,3)$ with ${s_\ell}=0$ and the
${\bf 6}=(2,0)$ with ${s_\ell}=1$. The other states are exotics.
When combined with the heavy quark $Q=c$, the ${\bf \bar 3}$ gives
the multiplet $\{\Xi_c^+,\ \Xi_c^0,\ \Lambda_c^+\,\}$ with spin-1/2,
and the ${\bf  6}$ gives the multiplets $\{\Sigma_c^{++},\
\Sigma_c^{+},\ \Sigma_c^{0},\ \Xi_c^{'+},\ \Xi_c^{'0},\ \Omega_c^0\,\}$
with spin-1/2 and
$\{\Sigma_c^{*++},\ \Sigma_c^{*+},\ \Sigma_c^{*0},\ \Xi_c^{'*+},\
\Xi_c^{'*0},\ \Omega_c^{*0}\,\}$ with spin-3/2.

To determine the energies of the states, we need to classify them according to
their values of $K$ using the allowed states given in Eq.~\possib. We will
discuss only the ${\bf\bar 3}=(0,1)$ and ${\bf 6}=(2,0)$ states here.
The state ${\bf\bar 3}=(0,1)$ with ${s_\ell}=0$ arises from using rule $(a)$
for the $K=0$, $3Y=2$ state in $\bar q$ and has energy $-3V_1/2$, and the state
${\bf\bar 3}=(0,1)$ with ${s_\ell}=1$ arises from using rule $(b)$ for the
$K=1$, $3Y=2$ state in $\bar q$ and has energy $V_1/2$. These states are
unique, and so must correspond to the states obtained by tensoring the ${\bf
8}=(1,1)$ soliton states with $\bar q$, \ie
\eqn\tbarstates{
\ket{{\bf\bar 3}\ 0;{\bf 8}}=\kket{{\bf\bar 3}\ 0}_0,\qquad
\ket{{\bf\bar 3}\ 1;{\bf 8}}=\kket{{\bf\bar 3}\ 1}_1.
}
The states ${\bf 6}=(2,0)$ with ${s_\ell}=2,1,0$ arises from using rule $(b)$
for the $K=1$, $3Y=2$ state in $\bar q$ and have energy $V_1/2$. A state ${\bf
6}=(2,0)$ with ${s_\ell}=1$ arises from using rule $(a)$ for the $K=0$, $3Y=2$
state in $\bar q$ and has energy $-3V_1/2$. The ${\bf 6}=(2,0)$ states with
${s_\ell}=1$, $\kket{{\bf 6} 1}_{0,1}$ must be linear combinations of the
states $\ket{{\bf 6}\ 1; {\bf 8}}$ and  $\ket{{\bf 6}\ 1; {\bf 10}}$
obtained by tensoring ${\bf 8}=(1,1)$ and ${\bf\bar{10}}=(3,0)$ with $\bar q$
respectively. We need to find the transformation relating the two sets of basis
states (analogous to Eq.~\twostates) to determine the energies of the ${\bf 6}$
states including the ${\bf 10}-{\bf 8}$ mass splitting.

The straightforward way to determine the linear combinations is to compute the
interaction hamiltonian for $SU(3)$. The result is essentially identical to
Eq.~\hsixj\ with the $SU(2)$ $6j$-symbol for isospin replaced by the
corresponding $SU(3)$ $6j$-symbol, and with the adjoint representation 1 of
$SU(2)$ replaced by the adjoint representation ${\bf 8}$ of $SU(3)$. The labels
$R,R'$ in the $SU(2)$ $6j$-symbol for spin refer to the isospin of the states
in the $SU(3)$ representation that satisfy the hypercharge constraint. The
$SU(3)$ $6j$-symbols are complicated, and have four additional labels because
there can be more than one singlet in the tensor product of three $SU(3)$
representations.

Instead of doing this, we will determine the linear combinations for the ${\bf
6}$ by a different method. Let us consider the state $\kket{{\bf 6} \ 1}_0$
which can be written as the linear combination
\eqn\lincom{
\kket{{\bf 6}\ 1}_0=\alpha \ket{{\bf 6}\ 1; {\bf 8}} + \beta \ket{{\bf 6} \ 1;
{\bf 10}}.
}
The ratio $\alpha/\beta$ is determined by making the right hand side an
eigenstate of $K$ with eigenvalue zero. The algebra is relegated to Appendix~A.
The result is that we get the same linear combination for $SU(3)$ as we did for
$SU(2)$ in Eq.~\twostates. Thus the hamiltonian Eq.~\twomod\ is the hamiltonian
for the ${\bf 6}$ of $SU(3)$, where $\Delta M$ is now the ${\bf 10}-{\bf 8}$
mass difference, and the mass formula Eq.~\callkleb\ now gives
the ${\bf 6}-{\bf \bar 3}$ mass difference.

The baryon-meson couplings in $SU(3)$ are easily determined from the results of
Sec.~6;  the only difference is that we replace Eqs.~\gai\ and \gav\ by
the corresponding $SU(3)$ expressions
\eqn\repgai{\eqalign{
&\phantom{\kket{}}_{0}\!\left.\bbra{R'\,a'\ {s_\ell}'\,m'}\right|
\Tr A\tau^r A^{-1}\tau^s\kket{R\,a\ {s_\ell}\,m}_0=\cr
\noalign{\smallskip}
&\qquad\qquad=(-2) \sqrt {\dim R\over \dim R'}
\clebsch {{\bf R}}{{\bf 8}}{{\bf R'}}as{a'}
\clebsch{{\bf R}}{{\bf 8}}{{\bf R'}}{b}{r}{b'},\cr
}}
and
\eqn\repgav{\eqalign{
&{}_{0}\!\left.\bbra{R'\,a'\ {s_\ell}'\,m'}\right|
S^r I^s\kket{I\,a\ {s_\ell}\,m}_0=
{1\over 4}\sqrt{\dim R\over \dim R'}\clebsch
{{\bf R}}{{\bf 8}}{{\bf R'}}as{a'}
\clebsch{{\bf R}}{{\bf 8}}{{\bf R'}}{b}{r}{b'},\cr
}}
where the states
$b$ and $b'$ have $I=s_\ell\ (s_\ell')$, $I_3=m\ (m')$, and $3Y=2$,
the factors on the right are $SU(3)$ Clebsch-Gordan coefficients,
and the dimensions are those of the $SU(3)$ representations.
The Clebsch-Gordan coefficients are evaluated in terms of $SU(2)$
Clebsch-Gordan coefficients
and the appropriate isoscalar factors.  Using the tensor methods
discussed in Appendix A, we find the isoscalar factors
\eqn\isos{\eqalign{
&\clebsch{{\bf 6}}{{\bf 8}}{{\bf 6}}{\Sigma_c}{\pi}{\Sigma_c}=
\sqrt{{3\over 5}},\cr
\noalign{\smallskip}
&\clebsch{{\bf 6}}{{\bf 8}}{{\bf \bar 3}}{\Sigma_c}{\pi}{\Lambda_c}=
{\sqrt{3} \over 2}. }}
This gives
\eqn\newgs{\eqalign{&g_2=-{9\over 7}\, g_A + {3\over 10}\, g,\cr
\noalign{\smallskip}
&g_3={15\over 14}\, g_A - {1\over 4}\, g,}}
compared to the $SU(2)$ prediction \gavii.\foot{Note that the relation
between $D$ and $g_A$ given in Eq.~\nucax\ is also modified by $SU(3)$
isoscalar factors.}
The ratio of the couplings is
\eqn\newratio{ {g_2\over g_3}=-{6\over5},}
for $SU(3)$, and differs from the $SU(2)$ prediction of Eq.~\gaviii\ by
the factor $\sqrt{24/25}=0.98$. The Skyrme model predictions for the
couplings are different for $SU(2)$ and $SU(3)$. This is a generic
feature which is a peculiarity of the Skyrme model. It arises because
the $SU(3)$ Clebsch-Gordan coefficients used are those for the baryon
representations for $N_c=3$ rather than those for the true large $N_c$
baryon representations which are Young tableaux with $N_c$ boxes \am.
If we had used the true large $N_c$ Clebsch-Gordan coefficients
(and retained only the leading term as $N_c\rightarrow\infty$),
we would have obtained the same results for $SU(2)$ and for $SU(3)$.
A simple way to see this is to note that the quark model results in the
large $N_c$ limit have the same group theoretic structure as the Skyrme
model \am, and that the quark model results for baryons with no strange
quark are obviously the same for $SU(2)$ and $SU(3)$.

\bigskip
\centerline{\bf Acknowledgements}
We would like to thank E.~Jenkins and M.~B.~Wise for
useful discussions. A.M. would like to thank the Fermilab theory group for
hospitality while this paper was being written.
This work was supported in part by DOE grant \doe,
and by a NSF Presidential Young Investigator award \pyiam.
\bigskip

\appendix{A}{Determining $K$ Eigenstates for $SU(3)$}
The state
\eqn\alincom{
\kket{{\bf 6}\ 1}_0=\alpha \ket{{\bf 6}\ 1; {\bf 8}} + \beta \ket{{\bf 6} \ 1;
{\bf 10}},
}
discussed in Eq.~\lincom\ was defined to be an eigenstate of $K$ with
eigenvalue 0. In this appendix, we determine the ratio of $\alpha/\beta$ for
which this is true. The state can be written as
\eqn\aexp{\eqalign{
\kket{{\bf 6}\, c\ 1\, m}_0=&\ \alpha \clebsch{{\bf 8}}{{\bf\bar3}}{{\bf 6}}abc
\clebsch{\half}{\half}{1}npm \sket{{\bf 8}\,a\ \half\,n}\hket{b\,p}\cr
\noalign{\smallskip}
&+ \beta \clebsch{{\bf 10}}{{\bf\bar3}}{{\bf 6}}{a'}{b'}c
\smallclebsch{\frac32}{\half}{1}{n'}{p'}m \sket{{\bf 10}\,a'\
\frac32\,n'}\hket{b'\,p'},
}}
in terms of the soliton states $\sket{\ }$ and the $\bar q$ states $\hket{\ }$.
The normalised soliton states are
\eqn\nsol{
\sket{R\,a\ {s_\ell}\,m} = \sqrt{\dim R}\ (-1)^{{s_\ell}+m}\int_{SU(3)}dg\
D^{*(R)}_{ad}(g)\ \hat U(g) \sket{\Sigma_0}
}
where the index $d$ is a state with $I={s_\ell}$, $I_3=-m$ and $Y=1$ \am. The
eigenvalues of $K$ are defined for the $\bar q$ state when the soliton is in
the state $\sket{\Sigma_0}$. Thus we only need to consider the soliton
wavefunction Eq.~\nsol\ when $g=1$, so that $D^{*(R)}_{ad}(g)=\delta_{ad}$.
With this substitution, one obtains
\eqn\aai{\eqalign{
&\kket{{\bf 6}\ 1}_0\rightarrow\ \alpha\sqrt8\ (-1)^{1/2+n} \clebsch{{\bf
8}}{{\bf\bar3}}{{\bf 6}}{K,\!-n}bc \clebsch{\half}{\half}{1}npm
\sket{\Sigma_0}\hket{b\,p}\cr
\noalign{\smallskip}
&\quad+\beta\sqrt{10}\ (-1)^{3/2+n'} \clebsch{{\bf 10}}{{\bf\bar3}}{{\bf
6}}{\Delta,\!-n'}{b'}c \smallclebsch{\frac32}{\half}{1}{n'}{p'}m
\sket{\Sigma_0}\hket{b'\,p'},
}}
where $\delta_{ad}$ has restricted the sum over $a$ to states with $Y=1$ and
$I=1/2$, and the sum over $a'$ to states with $Y=1$ and $I=3/2$.
The ${\bf\bar3}$ decomposes under $SU(2)$ into a doublet and a singlet, which
we will call $\bar q$ and $\bar s$ respectively. The ${\bf 6}$ decomposes into
the representations $I=1,1/2,0$ which will be called $qq$, $qs$ and $ss$
respectively. The $SU(3)$ Clebsch-Gordan coefficients in Eq.~\aai\ can be
written as the product of $SU(3)$ iso-scalar factors times $SU(2)$
Clebsch-Gordan coefficients. The non-zero isoscalar factors are $K\otimes \bar
q\rightarrow qq$ and $\Delta\otimes\bar q\rightarrow qq$. This restricts the
index $b,b'$ to $I=1/2$, $3Y=-1$. Since the index $b,b'$ cannot have $I=0$, we
see that there is no amplitude for the right hand side of Eq.~\aai\ to be in a
$K=1/2$ sector. We only need to ensure that it is orthogonal to the $K=1$
sector.
Rewriting the $SU(3)$ Clebsch-Gordan coefficients in terms of isoscalar
factors, we get
\eqn\aaii{\eqalign{
&\kket{{\bf 6}\ 1}_0\rightarrow\ \alpha\sqrt8\ (-1)^{1/2+n} \clebsch{{\bf
8}}{{\bf\bar3}}{{\bf 6}}K{\bar q}{qq}\clebsch{\half}{\half}1{-n}rc
\cr
&\qquad\qquad\qquad\qquad
\times\clebsch{\half}{\half}{1}npm \sket{\Sigma_0}\hket{r\,p}
\cr
\noalign{\smallskip}
&+ \beta\sqrt{10}\ (-1)^{3/2+n'} \clebsch{{\bf 10}}{{\bf\bar3}}{{\bf
6}}{\Delta}{\bar q}{qq}\smallclebsch{\frac32}{\half}{1}{-n'}{r'}{c} \cr
&\qquad\qquad\qquad\qquad\times\smallclebsch{\frac32}{\half}{1}{n'}{p'}m
\sket{\Sigma_0}\hket{r'\,p'}
}}
where $r$ and $r'$ are summed over 1,2. We rewrite the $H$ states as linear
combinations of $K$ states,
\eqn\aiii{
\hket{r\,p} = \clebsch{\half}{\half}{K}rpk \hket{K\,k}.
}
To ensure that the right hand side of Eq.~\aaii\ is orthogonal to $K=1$, the
coefficient of $\hket{K=1\,k}$ must vanish. This requires that
\eqn\aiv{\eqalign{
0=&\ \alpha\sqrt8\ (-1)^{1/2+n} \clebsch{{\bf 8}}{{\bf\bar3}}{{\bf 6}}K{\bar
q}{qq}\clebsch{\half}{\half}1{-n}rc
\clebsch{\half}{\half}{1}npm
\cr
\noalign{\smallskip}
&\qquad\qquad\qquad\times\sket{\Sigma_0} \clebsch{\half}{\half}{1}rpk
\cr
\noalign{\medskip}
&+ \beta\sqrt{10}\ (-1)^{3/2+n'} \clebsch{{\bf 10}}{{\bf\bar3}}{{\bf
6}}{\Delta}{\bar q}{qq}\smallclebsch{\frac32}{\half}{1}{-n'}{r'}{c}
\smallclebsch{\frac32}{\half}{1}{n'}{p'}m
\cr
\noalign{\smallskip}
&\qquad\qquad\qquad\times\sket{\Sigma_0}
\smallclebsch{\half}{\half}{1}{r'}{p'}k.
}}
The $SU(2)$ Clebsch-Gordan coefficients on the right hand side can be evaluated
in terms of $6j$-symbols. A simpler procedure is to pick the particular values
$k=1$, $m=1$ and $c=0$, for which the sums are trivial, and gives
\eqn\av{
0=\alpha\sqrt8\ \clebsch{{\bf 8}}{{\bf\bar3}}{{\bf 6}}K{\bar q}{qq}
- {1\over 2}\beta\sqrt{10}\ \clebsch{{\bf 10}}{{\bf\bar3}}{{\bf
6}}{\Delta}{\bar q}{qq}.
}

The isoscalar factors can be easily evaluated by tensor methods. The
Clebsch-Gordan coefficients for ${\bf 8}\otimes{\bf \bar3}\rightarrow{\bf 6}$
can be written as the $SU(3)$ invariant combination
\eqn\avi{
\chi\ {\bf 8}^k_j\  {\bf \bar3}_m \ {\bf 6}^\dagger_{kl}\ \epsilon^{jml}
}
in an obvious notation. To determine the normalisation constant $\chi$, we can
analyze the Clebsch-Gordan coefficients for ${\bf 6}^{11}$, for which Eq.~\avi\
gives
\eqn\avii{
\chi\ \left({\bf 8}^1_2\ {\bf \bar3}_3 - {\bf 8}^1_3\ {\bf \bar3}_2\right)
{\bf 6}^\dagger_{11},
}
which determines $\chi=1/\sqrt2$ for an amplitude normalised to unity. The
$SU(3)$ Clebsch-Gordan coefficient
\eqn\aviii{
\clebsch{\bf 8}{\bf \bar3}{\bf 6}{K^+}{\bar d}{uu}=
\clebsch{\bf 8}{\bf \bar3}{\bf 6}{K}{\bar q}{qq}
\smallclebsch{\half}{\half}{1}{\half}{\half}{1}
=-{1\over \sqrt2},
}
so that the isoscalar factor is
\eqn\avix{
\clebsch{\bf 8}{\bf \bar3}{\bf 6}{K}{\bar q}{qq}=-{1\over \sqrt2}.
}
Similarly, one determines the isoscalar factor
\eqn\ax{
\clebsch{\bf 10}{\bf \bar3}{\bf 6}{\Delta}{\bar q}{qq}={2\over \sqrt5},
}
using the normalised invariant combination
\eqn\axi{
\sqrt{3\over 5}\  {\bf 10}^{ijk}\ {\bf \bar3}_i\ {\bf 6}^\dagger_{jk}
}
Substituting in Eq.~\av, we find that $\beta=\sqrt2\alpha$, which is precisely
the linear combination obtained in eq.~\twostates\ for the state with $K=0$.
The orthogonal linear combination with $\alpha=-\sqrt2\beta$ must be the $K=1$
state, since that is the only other $\kket{{\bf 6}\ 1}_K$ state in the
spectrum.

\listrefs
\end